\documentclass[12pt]{article}
\usepackage{array, booktabs}
\usepackage{bm}
\usepackage{graphicx}
\usepackage{caption}
\usepackage{amsmath,amssymb,wasysym}
\usepackage{algorithm}
\usepackage{multirow}
\usepackage{algpseudocode}
\usepackage{lineno}  
\usepackage{setspace}  

\usepackage[utf8]{inputenc}
\usepackage{xcolor}
\usepackage{hyperref}
\usepackage{cleveref}
\usepackage[margin=1 in]{geometry}

\usepackage[utf8]{inputenc}
\usepackage[T1]{fontenc}
\usepackage{helvet}

\usepackage[affil-it]{authblk}
\usepackage{lineno}  
\usepackage{setspace}  

\makeatletter
\newcommand*{\centerfloat}{%
  \parindent \z@
  \leftskip \z@ \@plus 1fil \@minus \textwidth
  \rightskip\leftskip
  \parfillskip \z@skip}
\makeatother

\usepackage{xcolor}

\usepackage[
    backend=biber,natbib,
    style=nature,
]{biblatex}
\begin{document}

\title{Optimal Feeding of  \\ Swimming and Attached Ciliates}
\author{Jingyi Liu$^1$, Yi Man$^2$, John H. Costello$^{3,4}$, Eva Kanso$^{1,5}$\footnote{kanso@usc.edu} \\ 
\footnotesize{$^1$Department of Aerospace and Mechanical Engineering,  \\ University of Southern California, Los Angeles, California, USA} \\
\footnotesize{$^2$Mechanics and Engineering Science, Peking University, Beijing 100871, China} \\
\footnotesize{$^3$Department of Biology, Providence College, Providence, USA} \\
\footnotesize{$^4$Whitman Center, Marine Biological Laboratories, Woods Hole, USA}\\
\footnotesize{$^5$Department of Physics and Astronomy, University of Southern California, Los Angeles, California, USA}}


\date{\today}


\maketitle

Ciliated microorganisms near the base of the aquatic food chain either swim to encounter prey or attach at a substrate and generate feeding currents to capture passing particles. Here, we represent attached and swimming ciliates using a popular spherical model in viscous fluid with slip surface velocity that afford analytical expressions of ciliary flows. We solve an advection-diffusion equation for the concentration of dissolved nutrients, where the P'eclet number (Pe) reflects the ratio of diffusive to advective time scales. For a fixed hydrodynamic power expenditure, we ask what ciliary surface velocities maximize nutrient flux at the microorganism's surface. We find that surface motions that optimize feeding depend on Pe. For freely swimming microorganisms at finite Pe, it is optimal to swim by employing a  "treadmill" surface motion, but in the limit of large Pe, there is no difference between this treadmill solution and a symmetric dipolar surface velocity that keeps the organism stationary. For attached microorganisms, the treadmill solution is optimal for feeding at Pe below a critical value, but at larger Pe values, the dipolar surface motion is optimal. We verified these results in open-loop numerical simulations, asymptotic analysis, and using an adjoint-based optimization method. Our findings challenge existing claims that optimal feeding is optimal swimming across all P'eclet numbers, and provide new insights into the prevalence of both attached and swimming solutions in oceanic microorganisms.   

\spacing{1.5}


\section{Introduction}

Feeding of oceanic microbes is essential for their biological fitness and ecological function~\cite{solari2006multicellularity, shekhar2023cooperative, guasto2012fluid, gasol2018microbial, elgeti2015physics}. The metabolic processes of microbes, from small bacteria to larger ciliates like the Stentor or Paramecium, hinge on the absorption of particles or molecules at their surface~\cite{doelle2014bacterial, bialek2012biophysics, verni1997feeding,wan2020, pettitt2002, vopel2002}. These particles or molecules vary widely depending on the organism and encompass dissolved oxygen and other gases, lightweight molecules, complex proteins, organic compounds, and small particles and bacteria. The typical random motion of these particles and bacteria is akin to a diffusive process at the scale of the microorganism~\cite{magar2003, magar2005, michelin2011, berg2018, berg1977physics}. For simplicity, all cases will be collectively referred to as "nutrients." 

Ciliated microorganisms use surface cilia to generate flows in viscous fluids~\cite{emlet1990flow, christensen2003, kirkegaard2016filter, pettitt2002, sladecek1981, hartmann2007}. Ciliates either swim~\cite{michelin2010, michelin2011, andersen2020, bullington1930, guasto2012fluid} or attach to a surface and generate feeding currents~\cite{sleigh1976, pepper2010, pepper2013, wan2020, zima2009, vopel2002, andersen2020}. 
Whether motile or sessile, these ciliates perform work against the surrounding fluid, creating flow fields that affect the transport of nutrients and maintain a sufficient turnover rate of nutrients, unattainable by diffusion only~\cite{solari2006multicellularity}. These nutrients can thus be modeled using a continuous concentration field subject to diffusion and advection by the microorganism's induced flows. 

The coupling between diffusive and advective transport can be essential for microorganisms to achieve feeding rates that match their metabolic needs~\cite{solari2006multicellularity}. The relative importance of advective transport is quantified by the P\'{e}clet number Pe = $\tau_{\rm diff}/\tau_{\rm adv}$, defined as the ratio of diffusive $\tau_{\rm diff}$ to advective $\tau_{\rm adv}$ time scales. The diffusive time scale $\tau_{\rm diff} = a^2/D$ is given by the typical size of the organism $a$ and the diffusivity $D$ of the nutrient of interest, while the advective time scale $\tau_{\rm adv}= a/\mathcal{U}$ is governed by the flow speed $\mathcal{U}$ created by the microorganism.

To generate flows in a viscous fluid, a ciliated microorganism, through ciliary activity, must execute a series of irreversible surface deformations \cite{purcell1977life, lauga2009}. We call "stroke" such sequence of surface deformations. A stroke can induce a net force on the organism causing it to swim, or in the case of an attached organism, can require a reaction force, applied via a tether or a stalk, to resist swimming. Alternatively, the stroke itself could produce zero net force and be non-swimming. The question is, for a fixed rate of energy dissipation in the fluid, what are the optimal strokes that maximize nutrient flux at the organism's surface?

For a freely-moving organism, ~\cite{michelin2011} showed that the ``treadmill'' stroke, where on average all cilia exert tangential forces pointing from one end of the organism to the opposite end, is the only stroke that causes swimming. Importantly,~\cite{michelin2011} proposed that the treadmill stroke optimizes swimming and feeding at once for all Pe values. All other strokes were deemed suboptimal for feeding. 

In this study, we evaluate, given a fixed amount of available energy, the effect of surface velocities on feeding rates in attached ciliated microorganisms and it's comparison to swimming ciliated microorganisms. We consider a simplified spherical geometry, with the ciliated envelope modeled via a tangential slip velocity at the spherical surface~\cite{blake1971}. The Stokes equations is solved analytically using a linear decomposition of the ciliary stroke in terms of swimming and non-swimming modes~\cite{blake1971,magar2003,michelin2011}, which we then optimize to maximize the organism's nutrient uptake for a given energetic cost.

Our results can be organized as follows: we first compare analytic solutions of the Stokes equations around the sessile and motile ciliated sphere. The difference in flow fields is fundamental and not reconcilable by a mere inertial transformation. We solve numerically the advection-diffusion equation around the sessile and motile sphere for a range of Pe values and for three distinct strokes: the treadmill mode, a symmetric dipolar mode, and a symmetric tripolar mode. Only the treadmill mode leads to swimming in the motile sphere case and requires a tethering force in the sessile case. By symmetry, higher order modes are stationary, thus identical in the motile and sessile cases. 
We find that nutrient uptake depends non-trivially on Pe values and, we successfully validate our numerical results by conducting an asymptotic analysis in the two limits of large and small Pe for the sessile sphere and comparing these asymptotic results to their counterparts in the motile case~\cite{magar2003,michelin2011}. 
We then turn to optimal feeding strokes in the sessile case and seek, for a given amount of energy, the optimal stroke (possibly combining multiple simpler strokes) that maximizes nutrient uptake. We find consistent results through an open loop search and an adjoint-based optimization method. 
We conclude by commenting on the implications of our findings to understanding biological diversity at the micron scale.



\section{Mathematical formulation}
\label{sec:ciliates}

\subsection{Fluid flows around sessile and motile ciliates }

\begin{figure*}[!h]
    \centering
    \includegraphics{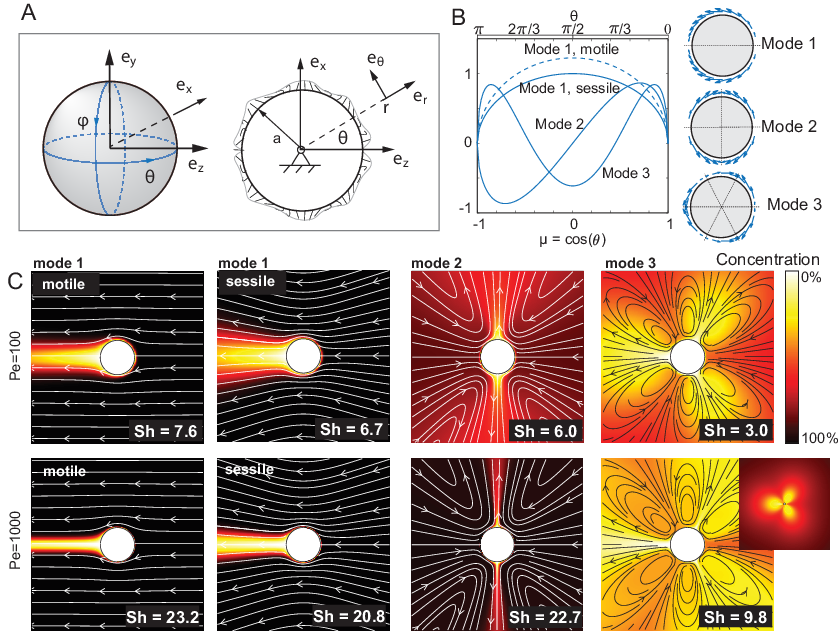}
    \caption{\textbf{Modeling motile and sessile ciliates at zero Reynolds number.} (A) Spherical envelope model with coordinates $(r, \theta, \phi)$, where $\theta \in [0,\pi]$ and, due to axisymmetry, $\phi \in [0,2\pi)$ is an ignorable coordinate. Ciliary motion is represented via a slip surface velocity. (B)  First three modes of surface velocity all at the same energy value: treadmill (mode 1), dipolar (mode 2), and tripolar (mode 3) modes, corresponding to $B_1 V_1(\mu)$ ($B_1=1$ for sessile and $\sqrt{3/2}$ for motile),  $B_2 V_2(\mu)$ and $B_3 V_3(\mu)$ with $B_2=\sqrt{3}$, $B_3=\sqrt{6}$ for both sessile and motile. Dotted lines represent lines of symmetry of surface velocity. (C) Flow streamlines (white) and concentration fields (colormap) at $\textrm{Pe}=100$ (top row) and $1000$ (bottom row) for the same hydrodynamic power $\mathcal{P}=1$ and distinct surface motions. In the treadmill mode, the streamlines, concentration field, and Sh number differ between the sessile and motile spheres, but are the same in the dipolar and tripolar surface modes.  }
    \label{fig:Coordinates}
\end{figure*}

The fluid velocity $\mathbf{u}$ in a three-dimensional domain bounded internally by a spherical ciliate of radius $a$ (Fig.\ref{fig:Coordinates}A) is governed by the incompressible Stokes equation \cite{kim1991}, 
\begin{align} \label{eq:stokes_eq}
    \begin{split}
    -\nabla p + \eta\nabla^2\mathbf{u} &= 0,\qquad 
    \nabla\cdot\mathbf{u} = 0,
\end{split}
\end{align}
where $p$ is the pressure field and $\eta$ is the dynamic viscosity.
To solve these equations, we consider the spherical coordinates $(r,\theta, \phi)$ and assume axisymmetric boundary conditions in $\phi$ at the spherical surface, with the axis of symmetry labeled by $z$ and the angle $\theta$ measured from the $z$-axis (Fig.\ref{fig:Coordinates}A). For notational convenience, we introduce the unit vectors $(\mathbf{e}_r,\mathbf{e}_\theta)$ and unit vector $\mathbf{e}_z$ along the axis of symmetry, $\mathbf{e}_z = \cos\theta\mathbf{e}_r - \sin\theta\mathbf{e}_\theta$ (Fig.\ref{fig:Coordinates}A).

Following Blake's envelope model~\cite{blake1971}, the cilia motion imposes a tangential slip velocity $\mathbf{u}(r=a,\theta)= V\mathbf{e}_\theta$ at the surface of the spherical boundary. We introduce the nonlinear transformation $\mu = \cos\theta$ and expand $V =\sum_{n=1}^\infty B_n V_n(\mu)$ in terms of the basis function $V_n(\mu)$ defined in terms of the Legendre polynomials $P_n(\mu)$ (\ref{append1}). 
All modes result in surface velocities with $\phi$-axis rotational symmetry. 
For mode 1, $B_1 = 1$ and $B_n = 0$ for all $n\neq 1$, the ciliary surface motion is referred to as a ``treadmill'' motion (Fig.~\ref{fig:Coordinates}B). 

We distinguish between two cases: a sessile sphere, fixed in space, and a motile sphere moving at a swimming speed $U$ in the $\mathbf{e}_z$ direction. The latter was considered in~\cite{blake1971,michelin2010,michelin2011}. In the motile case, the coordinate system $(r,\theta)$ is attached to the sphere and the equations of motion and boundary conditions are described in the body-fixed frame $(\mathbf{e}_r,\mathbf{e}_\theta)$. We get two sets of boundary conditions, 
\begin{equation} 
\begin{split} \label{eq:stokes_bc}
{\rm Sessile:}\quad \left. \mathbf{u} \right|_{r = a} = \sum_{n=1}^\infty B_n V_n(\mu) \mathbf{e}_\theta,  &\qquad \mathbf{u}|_{r\to\infty} = \mathbf{0},\\
{\rm Motile:}\quad  \left. \mathbf{u} \right|_{r = a} = \sum_{n=1}^\infty B_n V_n(\mu) \mathbf{e}_\theta,  &\qquad \mathbf{u}|_{r\to\infty} = -U\mathbf{e}_z.
\end{split}
\end{equation}
Substituting~\ref{eq:stokes_bc} into the general solution of \ref{eq:stokes_eq}, we obtain analytical expressions for the fluid velocity field and pressure field for sessile and motile sphere (Table~\ref{tab:stokes_solution}). 

In the motile case, we need an additional equation to solve for the swimming speed $U$. This equation comes from consideration of force balance. The hydrodynamic force acting on the sphere is given by $\mathbf{F}_h = \int \boldsymbol{\sigma}\cdot\mathbf{n}dS$, where $\boldsymbol{\sigma} = -p\mathbf{I} + \eta( \nabla\mathbf{u} + \nabla^T\mathbf{u})$ is the stress tensor. The total hydrodynamic force
$\mathbf{F}_h =0$ should be zero, leading to $U=2B_1/3$.

In the sessile case, the hydrodynamic force is balanced by an external force $\mathbf{F}_t$, provided by a tether or stalk, that fixes the sphere in space. From force balance,  $\mathbf{F}_t = -\mathbf{F}_h = 4\pi\eta a B_1\mathbf{e}_z$.

It is instructive to examine the leading order term in the fluid velocity $(u_r,u_\theta)$ in the sessile and motile case (Table~\ref{tab:stokes_solution}). In the sessile case, the far-field fluid velocity is of order $1/r$, similar to that of a force monopole  (Stokeslet). In the motile case, the far-field fluid velocity is of order $1/r^3$, as in the case of a three-dimensional potential dipole. 
The fluid velocity fields corresponding to the sessile and motile cases are not related to each other by a mere inertial transformation. 

\begin{table}[!t]
\caption{Comparison of Stokes flow around sessile and motile ciliate model. Mathematical expressions of fluid velocity field, pressure field, hydrodynamic power, forces acting on the sphere for both sessile and motile ciliated sphere. And the swimming speed for freely swimming ciliated sphere. All quantities are given in dimensional form in terms of the radial distance $r$ and angular variable $\mu=\cos\theta$.}
\begin{center}
\begin{tabular}{|l|l|}
\toprule
\multicolumn{2}{|c|}{\textbf{Sessile ciliated sphere}} \\ 
\toprule
Fluid velocity field & 
$u_{r}(r,\mu) =  \sum\limits_{n=1}^{\infty}\left(\dfrac{a^{n+2}}{r^{n+2}}-\dfrac{a^{n}}{r^{n}}\right)B_{n}P_{n}(\mu)$ \\ [2ex]
&
$u_{\theta}(r,\mu)  = \sum\limits_{n=1}^{\infty}\dfrac{1}{2}\left(\dfrac{na^{n+2}}{r^{n+2}}-\dfrac{(n-2)a^{n}}{r^{n}}\right)B_{n}V_{n}(\mu)$ \\[3ex]
Pressure field & 
$p(r,\mu) = p_{\infty} - \eta \sum\limits_{n=1}^{\infty}\dfrac{4n-2}{n+1}\dfrac{a^n}{r^{n+1}}B_{n}P_{n}(\mu)$ \\[3ex]
Energy dissipation rate & 
$\mathcal{P} = 16\pi a\eta\sum\limits_{n=1}^{\infty} \dfrac{B^2_n}{n(n+1)}$\\[3ex]
Hydrodynamic force & 
$\mathbf{F}_h = - 4\pi\eta a B_1\mathbf{e}_z$ 
\\[2ex]  
\toprule
\multicolumn{2}{|c|}{\textbf{Motile ciliated sphere}} 
    \\ \toprule
Fluid velocity field & 
$u_{r}(r,\mu) = \left(-\dfrac{2}{3} + \dfrac{2a^3}{3r^{3}}\right)B_{1}P_{1}(\mu)+\sum\limits_{n=2}^{\infty}\left(\dfrac{a^{n+2}}{r^{n+2}}-\dfrac{a^n}{r^{n}}\right)B_{n}P_{n}(\mu)$ \\[2ex]
 & 
$u_{\theta}(r,\mu) = \left(\dfrac{2}{3}+\dfrac{a^3}{3r^{3}}\right)B_{1}V_{1}(\mu) + \sum\limits_{n=2}^{\infty}\dfrac{1}{2}\left(\dfrac{n a^{n+2}}{r^{n+2}}-\dfrac{(n-2)a^n}{r^{n}}\right) B_{n}V_{n}(\mu)$ \\[3ex]
 Pressure field & 
$p(r,\mu) =p_{\infty} -  \eta\sum\limits_{n=2}^{\infty}\dfrac{4n-2}{n+1}\dfrac{a^n}{r^{n+1}}B_{n}P_{n}(\mu)$  \\[3ex]
 Energy dissipation rate & 
 $\mathcal{P} = 16\pi a\eta \left( \dfrac{1}{3}B_1^2 + \sum\limits_{n=2}^{\infty} \dfrac{B^2_n}{n(n+1)} \right)$ \\[3ex]
 Swimming speed & 
$U = \dfrac{2}{3}B_1$
\\[2ex]  
\toprule
\end{tabular}
\end{center}
\label{tab:stokes_solution}
\end{table}

In order to compare sessile and motile ciliates that exert the same hydrodynamic power $\mathcal{P}$ on the surrounding fluid, we introduce a characteristic velocity scale $\mathcal{U}$ based on the total 
 hydrodynamic power $\mathcal{U} = \sqrt{\mathcal{P}/(8\pi a\eta)}$. To obtain non-dimensional forms of the equations and boundary conditions, we consider the spherical radius $a=1$ and $\mathcal{T} = 1/\mathcal{U}=1$ as the characteristic length and time scales of the problem. These considerations impose the following constraints on the velocity coefficients ${B}_n$ (Table~\ref{tab:stokes_solution})
\begin{equation} \label{eq:tildeP_sum}
\begin{split}
     {\rm Sessile:}\quad\sum\limits_{n=1}^{\infty} \dfrac{2{B}^2_n}{n(n+1)}=1,\qquad
     {\rm Motile:}\quad \dfrac{2}{3}{B}_1^2 + \sum\limits_{n=2}^{\infty} \dfrac{2{B}^2_n}{n(n+1)}=1.
\end{split}
\end{equation}
Considering only the treadmill mode  leads to $B_1 = 1$ in the sessile case and $B_1 =\sqrt{3/2}$ in the motile case, with all other coefficients identically zero ($B_{n\neq 1}=0$). That is, for the same hydrodynamic power $\mathcal{P}$, the sessile sphere exhibits a slower surface velocity than the motile sphere ($B_1=1$ versus $B_1 =\sqrt{3/2}$). Considering only the second mode, we get $B_2 = \sqrt{3}$ and $B_{n\neq 2}=0$ in both the sessile and motile spheres, and when only the third mode is considered, $B_3=\sqrt{6}$ and $B_{n\neq 3}=0$ (Fig.~\ref{fig:Coordinates}B). In the sessile case, when a surface motion consists of multiple modes simultaneously, the portion of energy  assigned to each mode is denoted by $\beta_n^2$, such that $B_n  = \beta_n \sqrt{2/(n(n+1))}$. For example, if the total energy budget $\mathcal{P}$ is equally distributed between the first two modes, we have $\beta_1^2=0.5, \beta_2^2=0.5 $, and $B_1 = \sqrt{0.5}, B_2 = \sqrt{1.5}$.

\subsection{Advection-Diffusion Model of Nutrient Concentration} 
To determine the effect of the advective flows generated by the ciliated sphere on the nutrient concentration field around the sphere, we consider the advection-diffusion equation for the steady-state concentration $C$ of nutrients subject to zero concentration at the spherical surface~\cite{berg1977physics, bialek2012biophysics,magar2003,michelin2011}, 
\begin{equation}
    \mathbf{u}\cdot \nabla C = D \Delta C, \qquad  \left. C(\mu) \right|_{r=a=1} = 0,\quad  \left. C(\mu) \right|_{r\rightarrow\infty}= C_\infty.
    \label{eq:adv_diff}
\end{equation} 
We normalize the concentration field by its far-field value  $C_\infty$ at large distances away from the sphere and consider the transformation of variable $c = (C_\infty - C)/C_\infty$ \cite{magar2003,michelin2011}. Writing the advection-diffusion equation and boundary conditions~\eqref{eq:adv_diff} in non-dimensional form in terms of the new variable $c(r,\mu)$ yields
\begin{equation} 
    {\rm Pe}\; {\mathbf{u}}\cdot\nabla {c} = \Delta {c},\qquad \left. {c}(\mu)\right|_{r=a=1} = 1,\quad \left. {c}(\mu) \right|_{r\rightarrow\infty} = 0,
    \label{eq:adv_diff_nondim}
\end{equation}
where the P\'{e}clet number $\text{Pe}= a \mathcal{U}/D$, which quantifies the ratio of diffusive to advective time scales. When advection is dominant, the advective time is much smaller than the diffusive time and Pe $\gg 1$, when diffusion is dominant, the advective time is much larger and Pe $\ll 1$. At Pe $=1$, the two processes are in balance.

We substitute the analytical solutions of the flow field $\mathbf{u}$ from Table \ref{tab:stokes_solution} into~\eqref{eq:adv_diff_nondim}. We arrive at governing equations for the concentration field $c$, which we solve  analytically in the asymptotic limit of small and large P\'{e}clet numbers (~\ref{append2}) and numerically using a spectral method (~\ref{append3}).

\subsection{Sherwood number}

To quantify the uptake of nutrients at the surface of the sphere, we introduce the Sherwood number. 
The nutrient uptake rate is equal to the area integral of the concentration flux over the spherical surface 
\begin{equation}
I = -\oint \hat{\mathbf{n}}\cdot (- D \nabla C) \text{d}S,
\end{equation}
where $\text{d}S = 2 \pi a^2 \sin\theta \text{d}\theta$ is the element of surface area of the sphere. The sign convention is such that the concentration flux is positive if the sphere takes up nutrients. In the case of pure diffusion, the steady-state concentration obtained by solving the diffusion equation is given by  $C(r) = C_\infty(1 - a/r)$, and the steady-state inward current due to molecular diffusion is given by 
\begin{equation}
I_\textrm{diffusion} = 4 \pi a DC_\infty. 
\end{equation}
Accounting for both advective and diffusive transport,  the Sherwood number Sh is equivalent to a dimensionless nutrient uptake, where $I$ is scaled by $I_\textrm{diffusion}$,
\begin{align}
\begin{split} \label{eq:Sh}
    {\rm Sh} = \dfrac{I}{I_{\rm diffusion}} = 
    -\frac{1}{4\pi a D C_\infty}\oint \hat{\mathbf{n}} \cdot (-  D \nabla C) \text{d} S 
    = -\frac{1}{2}\int_{-1}^1 \left.\nabla c\cdot\mathbf{e}_r\right|_{r=a=1} d\mu.
    \end{split}
\end{align}

\section{Results}

\subsection{Comparison of feeding rates in sessile and motile ciliates}

In Fig.~\ref{fig:Coordinates}C, we show the streamlines (white)
around the motile and sessile spheres with slip surface velocity corresponding to the treadmill mode (mode 1), dipolar mode (mode 2), and tripolar mode (mode 3). In all cases, the hydrodynamic power $\mathcal{P}/8\pi\eta a = 1$ is held constant. Modes 2 and 3 produce zero net force on the sphere, causing no net motion even in the motile case, thus the streamlines are the same. Mode 1, the treadmill mode, is the only mode that leads to motility. The associated streamlines are shown in body-fixed frame in the motile case and in inertial frame in the sessile case.

The steady state concentration field (colormap) is obtained from numerically solving the advection-diffusion equation around the motile and sessile spheres at Pe = 100 and Pe = 1000. 
In the treadmill mode, the motile sphere swims into regions of higher concentration, which thins the diffusive boundary layer at its leading surface, leaving a trailing plume or “tail” of nutrient depletion. Similar concentration fields are obtained in the sessile sphere, albeit with wider trailing plumes, because, although the sphere is fixed, the surface treadmill velocity generates feeding currents that bring nutrients towards the surface of the sphere. 
In the dipolar and tripolar modes, feeding currents bring fresh nutrients to the spherical surface from, respectively, two opposite and three nearly-equiangular directions.

We evaluated the Sh number associated with each mode at both $\textrm{Pe} = 100$ and 
$\textrm{Pe}= 1000$. Clearly, larger $\textrm{Pe}$ leads to higher nutrient uptake. 
In the treadmill mode,
evaluating Sh number for the motile and sessile spheres led, respectively, to $7.6$ and $6.7$ at $\textrm{Pe} =100$ and $23.2$ and $20.8$ at $\textrm{Pe} =1000$.
Indeed, at each Pe, the motile sphere with treadmill surface velocity produced the largest Sh number, implying that for the same hydrodynamic power, motility maximized nutrient uptake. But the percent difference in nutrient uptake between the motile and sessile sphere decreased with increasing Pe, from $13.4\%$ to $11.5\%$. 
For the swimming sphere, comparing the treadmill mode and dipolar mode (mode 2), we found  $\textrm{Sh} = 7.6$ and $6.0$ at $\textrm{Pe}=100$ and $\textrm{Sh} = 23.2$ and $22.7$ at $\textrm{Pe}=1000$. That is,
for the motile sphere, the difference in nutrient uptake between mode 1 and mode 2 also decreased with increasing Pe from $26.7\%$ to $2.2\%$. 
Interestingly, for the sessile sphere, the nutrient uptake in mode 2 (Sh = 22.7) exceeded that of mode 1 (Sh = 20.8) at Pe = 1000, with a $9.1\%$ increase.

\begin{figure*}[!h]
    \centering
\includegraphics{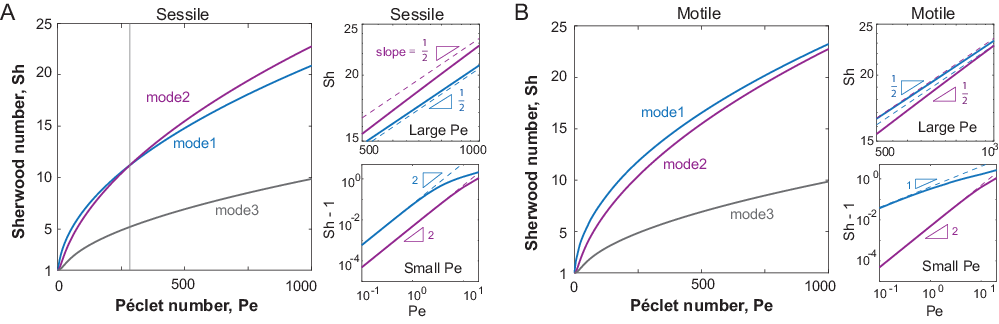}
    \caption{\textbf{Sherwood numebr as a function of P\'eclet number.} (A) sessile ciliate model  and (B) motile ciliate model  for the same hydrodynamic power $\mathcal{P}=1$. Solid lines are numerical calculations for mode 1 (blue), mode 2 (purple), and mode 3 (grey). Dashed lines and scaling laws in the limit of large and small Pe are obtained from asymptotic analysis for mode 1 (blue) and mode 2 (purple). }
    \label{fig: sh_1mode}
\end{figure*}

To probe the trends in Sh number over a larger range of Pe values, we computed, for the same hydrodynamic power, the steady state concentration field for the sessile and motile sphere, and for mode 1, 2, and 3, for $\textrm{Pe} \in [0,1000]$. 
The increment in Pe are dynamically adjusted from $\triangle{\rm Pe} = 0.01$ for Pe$<1$ (denser grid) to $\triangle{\rm Pe} = 100$ for ${\rm Pe} > 100$ (sparser grid), with $173$ discrete points in total between $\textrm{Pe} = 0$ and $\textrm{Pe} = 1000$. 
In Fig.\ref{fig: sh_1mode}A, we show the results for the sessile sphere. The solid lines in color blue, purple and grey represent mode 1, mode 2, and mode 3, respectively. Mode 1 exhibits the best feeding performance (highest Sh) for Pe $\leq 284$. Mode 2  exceeds mode 1 after this critical Pe value. At Pe$ = 1000$, Sh number of mode 2 is about $10\%$ higher than that of mode 1. Numerical results for the motile sphere are shown in Fig.\ref{fig: sh_1mode}B. Mode 1 outperforms modes 2 and 3 for the entire range of Pe values, but the difference between mode 1 and mode 2 seems to decrease and the two modes seem to approach each other asymptotically at larger Pe.

To further understand the asymptotic behavior of Sh in the limit of large Pe, we performed an asymptotic analysis to obtain the scaling of Sh with Pe
(\ref{append2}). To complete this analysis, we considered the two limits of small ${\rm Pe} \ll 1$ and large ${\rm Pe} \gg 1$ for the sessile ciliated sphere. Our approach is similar to that used in~\cite{magar2003,michelin2011} for the treadmill mode in the motile ciliated sphere. In Table \ref{tab:Sh_asymptotic}, we summarize the results of our asymptotic analysis for mode 1 and mode 2 of the sessile sphere. Note that the asymptotic analysis of mode 2 applies equally to the sessile and motile spheres.
For comparison reasons, we also include in this table the asymptotic results of \cite{magar2003, michelin2011} for the swimming ciliated sphere.
These asymptotic results are superimposed onto the numerical computations in the two insets in Fig.\ref{fig: sh_1mode}A and B.

At small Pe ($\textrm{Pe} \ll 1$), in the treadmill mode,  Sh scales with Pe$^2$ and Pe$^1$, respectively, for the sessile and motile spheres and, in the dipolar mode, Sh scales with Pe$^2$.
That is, at $\textrm{Pe}\ll 1$, the treadmill mode outperforms the dipolar mode in the motile case, and the motile sphere outperforms the sessile sphere in the treadmill mode.

At large Pe ($\textrm{Pe} \gg 1$), in the treadmill mode, Sh scales with $\sqrt{\rm Pe}$ in both the sessile and motile ciliated spheres. That is, there is no distinction in the scaling of Sh with Pe between the motile and sessile spheres. 
Interestingly, we found that in the dipolar mode, Sh also scales with $\sqrt{\rm Pe}$, indicating no distinction between the treadmill and dipolar modes.
The constant coefficients in the asymptotic scaling differ slightly: $\textrm{Sh} \approx0.65\sqrt{\rm Pe}$ (treadmill, sessile) and $\textrm{Sh} \approx0.72\sqrt{\rm Pe}$ (treadmill, motile), and $\textrm{Sh} \approx 0.74\sqrt{\rm Pe}$ (dipolar, both). Thus, in the motile sphere, the treadmill and dipolar modes perform nearly similarly in the large Pe limit, while in the sessile sphere, the dipolar mode outperforms the treadmill mode by a distinguishable difference (by about $10\%$).

\begin{table}[!t]
\caption{Asymptotic expressions for Sh as a function of Pe for sessile and swimming ciliated sphere model. The velocity coefficients associated with each mode are chosen satisfying same constraint hydrodynamic power.}
\begin{center}
\begin{tabular}{c|c|c|c|c|}
& \multicolumn{2}{c|}{\textbf{Large Pe limit}} & \multicolumn{2}{c|}{\textbf{Small Pe limit}} \\
\cline{2-5}
& \multicolumn{1}{c|}{\textbf{Sessile}} & \multicolumn{1}{c|}{\textbf{Motile} } & \multicolumn{1}{c|}{\textbf{Sessile}} & \multicolumn{1}{c|}{\textbf{Motile} } \\ 
\toprule
Mode 1 & 
$\textrm{Sh} = \dfrac{2}{\sqrt{3\pi}}{\rm Pe}^{\frac{1}{2}}$ &
$\textrm{Sh} = \dfrac{2}{\sqrt{3\pi}}\left(\dfrac{3}{2}\right)^{\frac{1}{4}}{\rm Pe}^{\frac{1}{2}}$, &
$\textrm{Sh} = 1 + \dfrac{43}{720}{\rm Pe}^2$ & 
${\rm Sh} = 1 + \dfrac{1}{3}\sqrt{\dfrac{3}{2}}{\rm Pe}$,\\[2ex] 
& &  \cite{magar2003,michelin2011} & & \cite{magar2003,michelin2011}\\[1ex] 
\hline
Mode 2 & \multicolumn{2}{c|}{$\textrm{Sh} = \dfrac{1}{\sqrt{\pi}}3^{\frac{1}{4}}{\rm Pe}^{\frac{1}{2}}$} & \multicolumn{2}{c|}{ $\textrm{Sh} = 1 + \dfrac{41}{8400}{\rm Pe}^2$ 
 } \\[2ex]
\hline 
\end{tabular}
\end{center}
\label{tab:Sh_asymptotic}
\end{table}

\subsection{Optimal feeding in sessile ciliates}

We next focused on the sessile ciliated sphere and, keeping the total hydrodynamic power constant, we investigated numerically how Sh number varies when multiple surface modes coexist. 

\begin{figure*}
    \centering
    \includegraphics{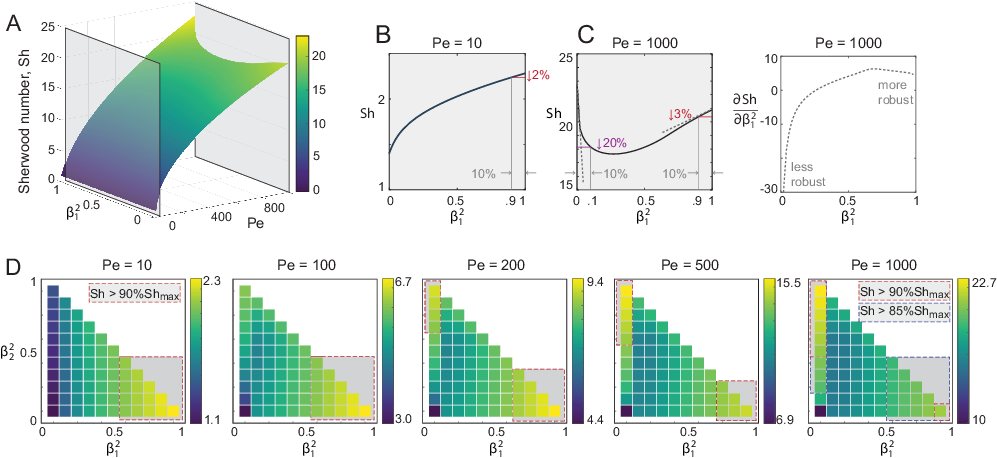}
    \caption{\textbf{Sherwood number for hybrid surface motions.} (A) Hybrid surface motions with two fundamental modes, treadmill and dipolar, and  constraint hydrodynamic power $\mathcal{P}/8\pi\eta a = \beta_1^2 +\beta_2^2 = 1$, where $\beta_1^2$ represents the portion of the energy assigned to mode 1. Sh versus Pe and $\beta_1^2$ as $\beta_1^2$ vary from $0$ to $1$. Close up at Sh number versus $\beta_1^2$ at (B)  Pe $=10$ and (C)  Pe $=1000$ (left), and  partial derivative of Sh with respect to $\beta_1^2$ (right).
    (D) Hybrid surface motions with three fundamental modes, treadmill, dipolar and tripolas, and constraint hydrodynamic power $\mathcal{P}/8\pi\eta a = \beta_1^2 +\beta_2^2 + \beta_3^2= 1$, where $\beta_1^2$ and $\beta_2^2$ represent, respectively, the portion of the energy assigned to the treadmill and dipolar modes.
    Colormap shows variation in Sh number as we vary the energy portions $\beta_1^2$ and  $\beta_2^2$ in the first two modes. Grey regions marked by red dash lines correspond to Sh within 10\% of corresponding maximal Sh. }
    \label{fig: sh_2mdoes}
\end{figure*}

In Fig.~\ref{fig: sh_2mdoes}A, we distributed the 
total hydrodynamic power into the first two modes only, assigning a fraction $\beta_1^2$ to mode 1 and the remaining fraction $1-\beta_1^2$ to mode 2. We varied $\beta_1^2$ from 1 (all hydrodynamic power assigned to mode 1) to 0 (all hydrodynamic power assigned to mode 2) at fixed intervals $\Delta \beta_1^2 = 0.05$. We also varied Pe from $0$ to $1000$ at $\Delta {\rm Pe} = 0.1$. For each combination of $({\rm Pe},\beta_1^2)$, we computed the steady state concentration field and calculated the resulting Sh number. We found that at small Pe, Sh increased monotonically as $\beta_1^2$ varied from 0 to 1, indicating that mode 1 is optimal. At larger Pe, a new local maximum appeared at $\beta_1^2=0$ (mode 2). This change is evident when comparing Fig.~\ref{fig: sh_2mdoes}B  and Fig.~\ref{fig: sh_2mdoes}C, which illustrate Sh as a function of $\beta_1^2$ at Pe = 10 and Pe = 1000, respectively. At Pe = 10, the maximal Sh at $\beta_1^2=1$ is a global optimum. At Pe = 1000, two local optima in Sh number are obtained at $\beta_1^2=1$ and $\beta_1^2=0$, with $\left.{\rm Sh}\right|_{\beta_1^2=0}>\left.{\rm Sh}\right|_{\beta_1^2=1}$, indicating that mode 2 is a global optimum. Interestingly, when calculating the sensitivity $\partial {\rm Sh}/\partial \beta_1^2$ of these maxima to variations in $\beta_1^2$, we found that the maximum at $\beta_1^2=0$ (mode 2) is more sensitive to variations in $\beta_1^2$, with $\left|\partial {\rm Sh}/\partial \beta_1^2 \right|_{\beta_1^2 = 0}\gg \left| \partial {\rm Sh}/\partial \beta_1^2 \right|_{\beta_1^2 = 1}$.
That is, a small variation in $\beta_1^2$ leads to larger drop in Sh at $\beta_1^2 =0$ (mode 2), while the same variation in $\beta_1^2$ leads to a small drop in Sh at $\beta_1^2 = 1$ (mode 1). In Fig.\ref{fig: sh_2mdoes}C, we show that a $10\%$ variation of $\beta_1^2$ leads to a $3\%$ drop from the optimal value at mode 1 and $20\%$ drop from optimal value at mode 2. 

We next considered the case when the total hydrodynamic power $\mathcal{P}$ is distributed over the first three modes, with a fraction $\beta_1^2$ assigned to mode 1, a fraction $\beta_2^2$ assigned to mode 2, and the remaining fraction $(1 - \beta_1^2 - \beta_2^2)$ assigned to mode 3.  We considered five values of Pe $=[10, 100, 200, 500, 1000]$. For each Pe value, we varied $\beta_1^2$ and $\beta_2^2$ from 0 to 1 at $\Delta \beta^2_{(\cdot)} = 0.1$, computed the steady state concentration field at each grid point and evaluated the corresponding Sh number.
Results are shown in Fig.~\ref{fig: sh_2mdoes}D. A similar trend appears: at small Pe, feeding is optimal when all the energy is assigned to mode 1 ($\beta_1^2 = 1, \beta_2^2 = 0$), but as Pe increases, a new local optimum appears at mode 2 ($\beta_1^2 = 0, \beta_2^2 = 1$). To test the sensitivity of these optima to variations in surface motion, we highlighted in light grey regions in the $(\beta_1^2,\beta_2^2)$ space that correspond to a 10\% drop in Sh number from the corresponding optimal value. Although at high Pe, mode 2 reaches higher values of Sh, it is more sensitive to variations in surface motion. The local optimum at mode 1 is more robust to such perturbations.

The optima in Fig.~\ref{fig: sh_2mdoes} are identified in an open loop search over the parameter spaces $(\beta_1^2, {\rm Pe})$ and $(\beta_1^2,\beta_2^2,{\rm Pe})$. Such open loop search becomes unfeasible when considering higher order surface modes. A closed loop optimization algorithm that seeks surface velocities that optimize Sh is needed. Here, we adapted the adjoint optimization method with gradient ascent algorithm used in~\cite{michelin2011} \ref{append4}. 

\begin{figure*}
    \centering
\includegraphics{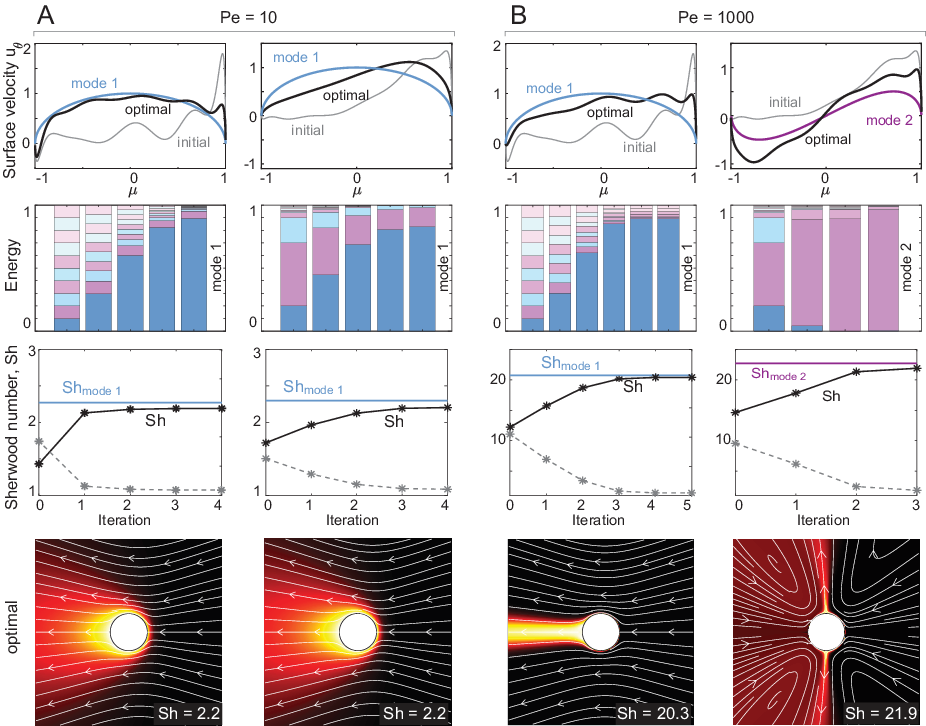}
    \caption{\textbf{Numerical optimization of surface motions that maximize feeding rates in sessile ciliates.} (A) Optimization results at Pe = $10$ for two different initial guesses.
    (B) Optimization results at Pe = $1000$ for the same two different initial guesses.
    Top row shows initial surface velocity (grey), optimal surface motion (black), and mode 1 (blue). Second row shows the energy distribution among ten different velocity modes at each iteration of the numerical optimization process. Third row shows Sherwood number (black) at each iteration, Sh number for only mode 1 Sh$_{\rm mode 1}$ (blue), and the difference in Sh number (Sh$_{\rm mode 1}$-Sh) (grey). Last row shows fluid and concentration fields under optimal surface conditions. In panel B, at Pe $=1000$, the numerical optimization algorithm converges to one of two distinct solutions, depending on initial conditions, that are either close to mode 1 (blue) or mode 2 (purple). }
    \label{fig: optimal_4case}
\end{figure*}

In Fig.~\ref{fig: optimal_4case},  we considered initial surface velocities with 10 modes, satisfying the constraint on the total hydrodynamic power $\sum_{n=1}^N \beta_n^2 = 1$. We used the closed loop optimization algorithm to identify optimal surface velocities that maximize Sh. 
In Fig.~\ref{fig: optimal_4case}A, we show the optimization results at Pe = 10 and two distinct initial conditions (grey line). The optimization algorithm converges to an optimal solution (black line) that is close to mode 1 (superimposed in blue). The energy distribution among all modes as a function of iteration steps shows that, while the initial energy was distributed among multiple modes, in the converged solution, energy is mostly assigned to mode 1. Indeed, comparing Sh number (black marker '$\ast$') to Sh number of mode 1 (blue line) shows that the optimal Sh converges to that of mode 1. Flow streamlines and concentration fields at these optima are shown in the bottom row of Fig.~\ref{fig: optimal_4case}A.

In Fig.~\ref{fig: optimal_4case}B, we show the optimization results at Pe = 1000 and the same two  initial conditions (grey line) considered in Fig.~\ref{fig: optimal_4case}A. Here, unlike in Fig.~\ref{fig: optimal_4case}A, the optimization algorithm converges to two different optimal solutions depending on initial conditions: one optimal (black line) is close to mode 1 (superimposed in blue)
and the other optimal is close to mode 2 (superimposed in purple), as reflected in the energy distribution (second row) and in Sh values (third row). Flow streamlines and concentration fields at these two distinct optima are shown in the bottom row of Fig.~\ref{fig: optimal_4case}B. The second optimal, the one corresponding to mode 2, exhibits higher Sh value. These results are consistent with the open loop analysis presented in Figs.~\ref{fig: sh_1mode} and~\ref{fig: sh_2mdoes}.

\section{Conclusion}

This work outlines several novel contributions. (1) we extended the envelope model to a fixed ciliated sphere, (2) analyzed the feeding rates around fixed and freely-swimming spheres and computed the Sherwood number numerically in a range of moderate Péclet values, (3) performed asymptotic analysis for Sh as a function of Pe in the two extreme (small and large) Pe limits and for the two first modes of surface velocities, and (4) computed optimal surface velocities that maximize feeding rates using an adjoint feedback optimization method. 

For motile ciliates, we found that assigning all energy into a treadmill surface velocity that induces swimming is an optimal strategy for maximizing feeding rate but only in a finite Pe range. This is in contrast to the findings in~\cite{michelin2011} that claimed optimality of the treadmill mode for all Pe.  In the limit of large Pe, we found no distinction in nutrient uptake between the treadmill mode and the symmetric dipolar mode that applies zero net force on the ciliated body, inducing no swimming motion.

For sessile ciliates, we found that the treadmill mode achieves optimal feeding rate at relatively low Pe values below a critical value Pe$_{\rm cr} \approx 280$, while the dipolar mode becomes optimal for Pe exceeding this threshold. Our asymptotic analysis supports that, in the large Pe limit, the dipolar surface mode outperforms other modes in terms of feeding rate. However, our sensitivity analysis shows that although at large Pe the treadmill mode leads to lower Sherwood  values, it is more robust to perturbations in the surface velocity. The dipolar mode leads to higher Sherwood values but it is significantly more sensitive to perturbations, with feeding efficiency dropping rapidly even with small perturbations in surface motion.

Our findings challenge previous assumptions that motility inherently improves feeding rate in ciliates~\cite{michelin2011,andersen2020}. We demonstrate that the optimal cilia-driven surface velocity for maximizing feeding rate varies significantly depending on the Péclet number, with distinct advantages observed for both motile and sessile ciliates under different conditions. This study enriches the understanding of the complexity of feeding strategies in ciliated microorganisms and highlights the importance of considering various environmental conditions when evaluating the ecological roles and evolutionary adaptations of these microbes.

\printbibliography






 \null
 \vfill
 \newpage

\renewcommand\thesection{APPENDIX \Alph{section}}
\setcounter{section}{0}


\section{General Solution of the Stokes Equations in Spherical Coordinates}
\label{append1}

\paragraph{Stokes equations.} For solving fluid velocity $\mathbf{u}$ at zero Re limit, the incompressible Stokes equation, we consider following approach. Due to the continuity property of the fluid, the velocity vector $\mathbf{u}$ can be expressed in terms of a vector potential $\Psi$ as $\mathbf{u}=\nabla\times\Psi$ \cite{batchelor1967}.
Taking the curl of the Stokes equation in~\eqref{eq:stokes_eq}, substituting $\mathbf{u}=\nabla\times\Psi$ into the resulting equation and using the incompressibility condition, we get the classic result that the vector potential $\Psi$ is governed by the bi-Laplacian $\nabla^2\cdot\nabla^2\Psi =0$~\cite{happel1965}.

To solve for the fluid velocity in the fluid domain bounded internally by a spherical boundary of radius $a$, it is convenient to introduce the spherical coordinates $(r,\theta,\phi)$ and associated unit vectors $(\mathbf{e}_r,\mathbf{e}_\theta,\mathbf{e}_\phi)$ (see Fig.~\ref{fig:Coordinates}.A). We express the fluid velocity in component form $\mathbf{u}\equiv (u_r,u_\theta,u_\phi)$. Here, we are interested only in axisymmetric flows, for which $u_\phi=0$ is identically zero, and the components of the vector potential $\Psi$ can be expressed in terms of an axisymmetric stream function $\psi$ (see, e.g., \cite{batchelor1967}), 
\begin{equation}
    \Psi=\left(0,0,\frac{\psi}{r\sin\theta}\right)^{T}.
    \label{eq:Psi}
\end{equation}
The non-trivial components $(u_r,u_\theta)$ of the fluid velocity are related to $\psi(r,\theta)$ as follows,
\begin{equation}
\begin{split} \label{eq:ur_uth}
    u_{r} =\frac{1}{r^{2}\sin\theta}\frac{\partial\psi}{\partial\theta},\qquad 
    u_{\theta} =-\frac{1}{r\sin\theta}\frac{\partial\psi}{\partial r}.   
\end{split}
\end{equation}
The streamfunction $\psi$ is governed by the biharmonic equation $E^2E^2(\psi)=0$ given in terms of the linear operator $E^2$,
\begin{equation} \label{eq:E_operator}
    E^{2}=\frac{\partial^{2}}{\partial r^{2}}+\frac{\sin\theta}{r^{2}}\frac{\partial}{\partial\theta}\left(\frac{1}{\sin\theta}\frac{\partial}{\partial\theta}\right). 
\end{equation}
This biharmonic equation $E^2E^2(\psi)=0$ can be solved analytically for arbitrary boundary conditions in terms of $r$ and the coordinate $\mu$ obtained via the nonlinear transformation of  coordinate $\mu = \cos\theta$. Explicitly expression of $\psi(r,\mu)$ can be found in~\cite{happel1965},
%
%
\begin{equation}
    \psi(r,\mu)=\sum_{n=2}^{\infty}(a_{n}r^{n}+b_{n}r^{-n+1}+c_{n}r^{n+2}+d_{n}r^{-n+3})F_{n}(\mu).
    \label{eq:psi}
\end{equation}
Here, $F_n(\mu)= -\int P_{n-1}(\mu)d\mu$ are solution functions related to the Legendre Polynomials of the first kind $P_{n}(\mu)$, satisfying equation $(1-\mu^{2})\frac{d^{2}F_n}{d\mu^{2}}+n(n-1)F_n = 0$, and  $a_{n}$, $b_{n}$, $c_n$, and $d_n$ are unknown coefficients.

By subtituting \ref{eq:psi} into \ref{eq:ur_uth}, we obtain the velocity components $(u_r,u_\theta)$,
\begin{equation}
\begin{split}
    u_{r}(r,\mu) &=\left(\alpha_0 + \alpha_{13}\frac{1}{r^3}+\alpha_{11}\frac{1}{r}\right)P_{1}(\mu) + \sum_{n=2}^\infty \left(\alpha_{n+2}\frac{1}{r^{n+2}} + \alpha_{n}\frac{1}{r^{n}}\right)P_{n}(\mu),\\
    u_{\theta}(r,\mu) &= \left( -\alpha_{0} + \alpha_{13}\frac{1}{2r^{3}} - \alpha_{11}\frac{1}{2r}\right)V_{1}(\mu) + \sum_{n=2}^\infty \frac{1}{2}\left(\alpha_{n+2}\frac{n}{r^{n+2}} + \alpha_{n}\frac{n-2}{r^{n}}\right)V_{n}(\mu),
   \label{eq:u_general}
\end{split}
\end{equation}
where $\alpha_0$, $\alpha_{11}, \ldots, \alpha_n$ are unknown coefficients, related to $a_n$, $b_n$, $c_n$, $d_n$ in~\eqref{eq:psi}, to be determined from boundary conditions. The basis functions $V_{n}(\mu)$ are defined as
\begin{equation}
    V_{n}(\mu) = - \frac{2}{\sqrt{1-\mu^2}}\int P_{n}(\mu)d\mu=\frac{2}{n(n+1)}\sqrt{1-\mu^2}P'_{n}(\mu),
\end{equation}
where $P_n'(\mu) = \frac{dP_n(\mu)}{d\mu}$. Both the Legendre polynomials $P_n(\mu)$ and basis functions $V_n(\mu)$ satisfy the orthogonality conditions
\begin{equation}
\begin{split}
    \int_{-1}^{+1}P_{n}(\mu)P_{m}(\mu)d\mu =\frac{2}{2n+1}\delta_{nm},\qquad \qquad
    \int_{-1}^{+1}V_{n}(\mu)V_{m}(\mu)d\mu = \frac{8}{n(n+1)(2n+1)}\delta_{mn}.
\end{split}
\end{equation}

\paragraph{Analytical expressions of the pressure field.}
The pressure is obtained by substituting~\eqref{eq:u_general} into Stokes equation \eqref{eq:stokes_eq} and integrating, 
yielding the pressure field $p(r,\mu)$ as
\begin{equation}
    p = p_{\infty} + \eta \sum_{n=1}^{\infty}\frac{4n-2}{n+1}\frac{\alpha_n}{r^{n+1}}P_{n}(\mu).
    \label{eq:p_general}
\end{equation}

\paragraph{Analytical expressions of the stress field.}
 The fluid stress tensor $\boldsymbol{\sigma}$ is given by $\boldsymbol{\sigma} = -p\mathbf{I} + \eta( \nabla\mathbf{u} + \nabla^T\mathbf{u})$. For the axisymmetric flows considered here, $\boldsymbol{\sigma}$ admits three non-trivial stress components 
\begin{align}
\begin{split}
    \sigma_{rr} = -p+2\eta\frac{\partial u_{r}}{\partial r},\qquad
    \sigma_{r\theta} = \eta\left(\frac{1}{r}\frac{\partial u_r}{\partial\theta}+r\frac{\partial}{\partial r}(\frac{u_{\theta}}{r})\right),\qquad
    \sigma_{\theta\theta} = -p+2\eta \left(\frac{1}{r}\frac{\partial u_\theta}{\partial\theta}+\frac{u_r}{r}\right).
    \label{eq:stress_def}
\end{split}
\end{align}
Explicit expressions for the stress components are obtained by substituting~\eqref{eq:u_general} and~\eqref{eq:p_general} into~\eqref{eq:stress_def}.

\paragraph{Hydrodynamic force acting on the sphere.} The hydrodynamic force exerted by the fluid on the sphere can be calculated by integrating  the stress tensor $\boldsymbol{\sigma}$ over the surface $S$ of the sphere. Due to axisymmetry, only the force in the direction of the axis of axisymmetry, taken to be the $z$-axis, is non-zero,
\begin{equation}
    \mathbf{F} = \int_S \boldsymbol{\sigma}\cdot \hat{\mathbf{n}}\textrm{d}S = -4\pi\eta \alpha_{11}\mathbf{e}_z.
    \label{eq:general_force}
\end{equation}

\paragraph{Viscous dissipation energy} The energy dissipation rate is defined as the volume integral over the entire fluid domain $V$ of the inner product of the velocity strain rate tensor $\mathbf{e} = \frac{1}{2}(\nabla\mathbf{u} + \nabla\mathbf{u}^{\rm T})$ and stress tensor $\boldsymbol{\sigma}$, which, given proper decay at infinity, can be expressed as an integral over the surface of the sphere by applying the divergence theorem (see, e.g.,~\cite{kim1991})
\begin{equation} \label{eq:energy_dissipation_formula}
    \mathcal{P} = \int_V \mathbf{e}:\boldsymbol{\sigma} \textrm{d}V =   \int_{S} \mathbf{u}\cdot(\boldsymbol{\sigma}\cdot\hat{\mathbf{n}}) \textrm{d}S.
\end{equation}

This general solution is applicable to moving or stationary sphere with slip or no slip boundary conditions. With prescribed boundary condition, we can obtain fluid solution to rigid sphere and haired sphere in uniform flow, which are consist with exist study. In this article, we apply the general solution approach to sessile haired sphere in still fluid.

\section{Asymptotic Analysis}
\label{append2}
In this section, we consider the flow field generated by the ciliary motion and derive an asymptotic solution of the Sherwood number for sessile ciliated sphere in the limit of large and small Péclet, respectively. Particulary, we seek asymptotic expressions associated with mode 1 and mode 2. 

\subsection{Large Pe limit}
Here, we start with a general expression of velocity field corresponding to the $n^{th}$ mode surface velocity at the unit sphere $a=1$, for which $\left. u_{\theta}(\mu)\right|_{r=1} = {B}_n V_n(\mu)$, where $n=1, n=2$ are considered later in this paper. The flow field corresponding to $n^{th}$ mode is given by (see Tables~\ref{tab:stokes_solution})
\begin{align}\label{eq:velocity_mode_n}
    u_{r} = \left(\frac{1}{r^{n+2}}-\frac{1}{r^{n}}\right)B_{n}P_{n}(\mu),\qquad
    u_{\theta} = \left(\frac{n}{r^{n+2}}-\frac{n-2}{r^{n}}\right)\frac{B_{n}}{2}V_n(\mu),
\end{align}
We take the Taylor series expansion of flow field at $r=1$ and  keep only the leading order terms,
\begin{equation}
    \begin{split}
    \begin{array}{llcll}
         u_{r}(r,\mu) &= \left. u_r\right|_{r=1} + \left. \frac{\partial u_r}{\partial r}\right|_{r=1}(r-1) +\dots = - 2B_n(r-1)P_n(\mu) + \dots, \\[1ex]
         u_{\theta}(r,\mu) &= \left. u_\theta\right|_{r=1} + \left.\frac{\partial u_\theta}{\partial r}\right|_{r=1}(r-1) + \dots 
         = B_nV_n(\mu) + \dots.
        \end{array}
    \end{split}
    \label{eq:u_linear}
\end{equation}
We define the temporary variable $y = r-1$ (not to confused with the $y$-coordinated in the inertial $(x,y,z)$ space). The region $y\ll 1$ represents a thin boundary layer around the spherical surface. Since the concentration boundary layer is expected to be thinner as Pe increases, we rescale $r-1 = y = {\rm Pe}^{-m} Y$, where $Y$ is a new variable. 

Next, we substitute the linearized flow field $\mathbf{u}$ from~\eqref{eq:u_linear} into the advection-diffusion equation~\eqref{eq:adv_diff} and use the new variable $r-1 = {\rm Pe}^{-m} Y$. Keeping only the leading-order terms, we obtain the advection operator $\mathbf{u}\cdot\nabla $,
\begin{align}
    u_{r}\frac{\partial}{\partial r} = - 2B_n P_n Y \frac{\partial}{\partial Y},\qquad
    u_{\theta}\frac{1}{r}\frac{\partial}{\partial \theta}  = -B_nV_n\sqrt{1-\mu^2}\frac{1}{{\rm Pe}^{-m}Y+1}\frac{\partial}{\partial\mu}.
    \label{eq:adv_op_linear}
\end{align}
Similarly, rewriting the Laplacian operator using the new variable $r-1 = {\rm Pe}^{-m} Y$, we arrive at
\begin{align}
    \nabla^2 = {\rm Pe}^{2m} \frac{\partial^2}{\partial Y^2} + \frac{2{\rm Pe}^m}{{\rm Pe}^{-m}Y+1}\frac{\partial}{\partial Y} + \frac{1}{ ({\rm Pe}^{-m}Y+1)^2 }\frac{\partial}{\partial\mu}\left( (1-\mu^2)\frac{\partial}{\partial \mu}\right).
    \label{eq:diff_op_linear}
\end{align}
The leading-order term in the Laplacian operator scales with ${\rm Pe}^{2m}$, while the leading-order term in the advection operator scales with ${\rm Pe}^0$. Matching order on both sides of the advection-diffusion equation, we have $2m=1$. Thus, $m=\frac{1}{2}$, and we have $r-1 = {\rm Pe}^{-1/2}Y$. 

We now substitute~\eqref{eq:adv_op_linear} and~\eqref{eq:diff_op_linear}, with $m=1/2$ into the dimensionless advection-diffusion equation~\eqref{eq:adv_diff_nondim}, keeping only the leading-order terms, we arrive at
\begin{equation}
     -B_n\left( 2P_nY\frac{\partial c}{\partial Y} + V_n\sqrt{1-\mu^2}\frac{\partial c}{\partial\mu} \right) = \frac{\partial^2 c}{\partial Y^2}.
     \label{eq:c_linear_temp}
\end{equation}
We define a similarity variable $Z = Y/g(\mu)$ such that, by the chain rule,
\begin{equation}
    \frac{\partial}{\partial Y} =  \frac{1}{g}\frac{\partial}{\partial Z},\quad
    \frac{\partial}{\partial \mu} = g'\frac{\partial}{\partial g} = -\frac{Z g' }{g}\frac{\partial}{\partial Z}.
\end{equation}
We substitute into~\eqref{eq:c_linear_temp} and rearrange terms to arrive at the ordinary differential equation
\begin{equation}
     \frac{\partial^2 c}{\partial Z^2} + B_n\left( 2P_n g^2 - \frac{1}{n(n+1)}(1-\mu^2)P'_n gg' \right)Z\frac{\partial c}{\partial Z}=0.
\end{equation}
For a similarity solution to exist, the term $ 2\mu g^2 - (1-\mu^2)g g' $ needs to be equal to a constant. Setting the value of this constant to 2, the problem becomes that of solving
\begin{equation}
    \frac{d^2 c}{d Z^2} + 2Z\frac{d c}{dZ} =0, \qquad
    B_n \left( 2P_n g^2 - \frac{1}{n(n+1)}P'_n(1-\mu^2)(g^2)' \right) = 2.
\end{equation}
For the first mode, $n=1$ and $P_1 = \mu$, the solution is in the form,
\begin{equation}
    \begin{split}
        c(Z) = C_1 {\rm erf}(Z) + C_2, \qquad 
        g^2(\mu) = \frac{C_3 - 12\mu + 4\mu^3}{3B_1(\mu^2 - 1)^2}.
    \end{split}
\end{equation}
Here, $C_1$, $C_2$ and $C_3$ are unknown constants to be determined from the boundary conditions $c(Z=0)=1$, $c(Z\rightarrow\infty)=0$, and from the condition that at $\mu=1$, the concentration field and the function $g(\mu)$ must be bounded \cite{magar2003}. Put together, we get that $C_3 = 8$, $C_2 = 1$, and $C_1 = -1$. Thus, the asymptotic solution of the concentration field in the limit of large P\'eclet, Pe $\gg 1$, is given by 
\begin{align}
    \begin{split}
        c(Z) = 1-{\rm erf}(Z) = {\rm erfc}(Z),
    \end{split}
\end{align}
where
\begin{align}
    \begin{split}
        Z = \text{Pe}^{\frac{1}{2}} \dfrac{(r-1)}{g(\mu)},\qquad 
        g(\mu) = \sqrt{\frac{8 - 12\mu + 4\mu^3}{3B_1(1 - \mu^2)^2}} .
    \end{split}
\end{align}
The Sherwood number at large Pe is given by, 
\begin{equation}
\begin{split}
    {\rm Sh}_{\rm mode \;1} &= -\frac{1}{2}\int_{-1}^1 \left. \frac{\partial c}{\partial r}\right|_{r=1} d\mu = \frac{1}{2}{\rm Pe}^{\frac{1}{2}} \int_{-1}^1 \left. \frac{2}{\sqrt{\pi}} e^{-Z^2} \frac{1}{g}\right|_{r=1} d\mu  \\
    &= \frac{1 }{\sqrt{\pi}} {\rm Pe}^{\frac{1}{2}}\int_{-1}^1  \sqrt{\frac{3B_1(1 - \mu^2)^2}{8 - 12\mu + 4\mu^3}} d\mu
    = \sqrt{\frac{4B_1}{3\pi}}{\rm Pe}^{\frac{1}{2}},
\end{split}
\end{equation}
Similarly, for the second mode, $n=1$ and $P_2 = \frac{1}{2}(3\mu^2-1)$, we obtain the solution
\begin{align}
    \begin{split}
       Z = \text{Pe}^{\frac{1}{2}} \dfrac{(r-1)}{g(\mu)},\qquad 
        g(\mu) = \sqrt{\frac{1 + \mu^4 - 2\mu^2}{B_2(1 - \mu^2)^2\mu^2}} .
    \end{split}
\end{align}
The Sherwood number at large Pe is
\begin{equation}
\begin{split}
    {\rm Sh}_{\rm mode\; 2} = \sqrt{\frac{B_2}{\pi}}{\rm Pe}^{\frac{1}{2}},
\end{split}
\end{equation}

\subsection{Small Pe Limit}
At small Péclet number, we expand the concentration field as 
\begin{equation}
c = {\rm Pe}^0 c_0 + {\rm Pe}^1 c_1 + {\rm Pe}^2c_2 +\dots
\end{equation}
We substitute the expanded concentration into the dimensionless advection-diffusion equation~\eqref{eq:velocity_mode_n}, we arrive at the following system of equations, to be solved at each order in Pe, 
\begin{equation}
\begin{split}
& \text{Order} \ 0: \qquad \qquad \ 0 = \nabla^{2}c_{0},\quad \  \left. c_{0}\right|_{r=1} = 1,\quad \left. c_0\right|_{r\rightarrow\infty} = 0.\\
& \text{Order} \ 1:  \qquad \mathbf{u}\cdot \nabla c_0 = \nabla^{2}c_1,\quad \left.c_1\right|_{r=1} = 0,\quad \left.c_1\right|_{r\rightarrow\infty} = 0. \\
& \text{Order} \ 2: \qquad  \mathbf{u}\cdot \nabla c_1 = \nabla^{2}c_2,\quad \left.c_2\right|_{r=1} = 0,\quad \left.c_2\right|_{r\rightarrow\infty} = 0.
\end{split}
\label{eq:c_expand}
\end{equation}
At the leading order, the solution is simply $c_0 = 1/r$. To find the solution at higher orders, we substitute the velocity field~\eqref{eq:velocity_mode_n} into the higher order equations in~\eqref{eq:c_expand}. At order Pe${}^1$, we get
\begin{align}
    -B_n\left(\frac{1}{r^{n+2}}-\frac{1}{r^n}\right)\frac{1}{r^2}P_n(\mu) = \nabla^{2}c_1,\quad c_1(r=1) = 0.
\end{align}
Recall that the Legendre polynomials satisfy the Legendre differential equation $\frac{d}{d\mu}[(1-\mu^2)P_m'] + m(m+1)P_m = 0 $. Expanding $c_1$ in terms of Legendre polynomials $c_1(r,\mu) = \sum_{m=0}^{\infty} R_m(r)P_m(\mu)$, 
and substituting back into the above equation, we get
\begin{align}
    -B_n\left(\frac{1}{r^{n+2}}-\frac{1}{r^n}\right)P_n(\mu) = \sum_{m=0}^\infty \left(\frac{d }{d r}(r^2\frac{d R_{m}}{d r}) - m(m+1)R_{m} \right) P_m(\mu).
\end{align}
By equating the Legendre polynomials on both sides in the above equation, we get that only the term $R_n(r)$ survives,
\begin{align}
    -B_n\left(\frac{1}{r^{n+2}}-\frac{1}{r^n}\right) = r^2 R''_{1n} + 2r R'_{1n} - 6 R_{1n}.
\end{align}
For the first mode, solving the above ordinary differential equations with $n=1$, taking into consideration the boundary conditions, we get the solution for $c_1$,
\begin{equation}
    c_1(r,\mu) = \left( \frac{3}{4} r^{-2} - \frac{1}{4}r^{-3} - \frac{1}{2} r^{-1} \right)B_1 \mu. 
\end{equation}
Repeating the same procedure at order ${\rm Pe}^2$, we  get that
\begin{equation}
    \nabla^{2}c_2 = B_1^2\left(\frac{4 - 9r + 2r^2 + 3r^3 }{12 r^7} P_0(\mu) 
    - \frac{(r-1)(5 - 4r - 9r^2 + 6r^3)} {8 r^7}\frac{2}{3} P_2(\mu)\right).
\end{equation}
We expand $c_2 =\sum_{m=0}^{\infty} R_m(r)P_m(\mu)$ in terms of  a Legendre polynomial expansion  with unknown $R_m(r)$ and substitute back in the above equation; we get that only the terms $R_0(r)$ and $R_2(r)$ survive, such that the general solution for $c_2(r,\mu)$ is given by 
\begin{equation}
    c_2 (r,\mu)= R_0(r)P_0(\mu) + R_2(r)P_2(\mu).
\end{equation}
One can readily verify that the ordinary differential equation governing  $R_0(r)$ is given by
\begin{equation}
    r^2 R_0'' + 2r R_0' = B_1^2\frac{4 - 9r + 2r^2 + 3r^3 }{12 r^5}.
\end{equation}
The solution to this equation, taking into account the boundary conditions, is of the form
\begin{align}
    R_0(r) = B_1^2\left(- \frac{77}{720 r} + \frac{1}{60 r^5} - \frac{1}{16 r^4} + \frac{1}{36 r^3} + \frac{1}{8 r^2} \right).
\end{align}
It turns out that, for computing the Sherwood number below, only $R_0(r)$ is needed, as shown in 
\begin{align}
 -\frac{1}{2}\int_{-1}^{1}\left.\frac{\partial c_2}{\partial r}\right|_{r=1}d\mu = - \frac{1}{2}\sum_{m=0}^\infty \left. \frac{d R_m}{d r}\right|_{r=1}\int_{-1}^{1} P_m(\mu)d\mu = \left. \frac{d R_m}{d r}\right|_{r=1} \delta_{m0}
\end{align}
with implementing the property of Legendre polynomials $\int_{-1}^1 P_m(\mu)d\mu = 2\delta_{m0}$. We thus need not calculate $R_2(r)$. 
For the first mode, Sh in the small Pe limit is given by 
\begin{align}
\begin{split}
    {\rm Sh}_{\rm mode\; 1} &= -\frac{1}{2}\int_{-1}^{1}\left.\frac{\partial c_0}{\partial r}\right|_{r=1}d\mu  - \frac{1}{2}\left( \int_{-1}^{1} \left. \frac{\partial c_1}{\partial r}\right|_{r=1}d\mu \right) {\rm Pe} - \frac{1}{2}\left( \int_{-1}^{1} \left. \frac{\partial c_2}{\partial r}\right|_{r=1}d\mu \right) {\rm Pe}^2 \\
    &= 1 + \frac{43B_1^2}{720}{\rm Pe}^2.
\end{split}
\end{align}
Similarly, we can compute Sherwood number for surface velocity containing only the second mode by replacing velocity field with $n=2$,
\begin{align}
    {\rm Sh}_{\rm mode\; 2} = 1 + \frac{41B_2^2}{25200}{\rm Pe}^2.
\end{align}


\section{Spectral Method} 
\label{append3}
We solve~\eqref{eq:adv_diff_nondim} using the Legendre spectral method; see, e.g.,~\cite{michelin2011}. To this end, we first expand the concentration field $c(r,\mu)$ in terms of the Legendre polynomials $P_m(\mu)$
\begin{align}\label{eq:c_spectral}
    {c}(r,\mu) = \sum_{m=0}^\infty C_m(r)P_m(\mu),
\end{align}
where $C_m(r)$ are unknown coefficients associated with Legendre basis functions $P_m(\mu)$. We substitute \eqref{eq:c_spectral} into \eqref{eq:adv_diff_nondim} and project the governing equation onto the $k^{th}$ Legendre polynomial $P_k(\mu)$, to arrive at an infinite set of coupled boundary-valued ordinary differential equations for the unknown coefficients $C_k(r)$, $k=0,\ldots,\infty$
\begin{align}
\begin{split}
    {\rm Pe}\sum_{n=1}^{\infty}\sum_{m=0}^{\infty}{B}_{n} \left(E_{mnk}f_{nr}\frac{\partial C_{m}}{\partial r} - F_{mnk}f_{n\theta} \frac{C_{m}}{r}\right)
    =\frac{\partial^{2}C_{k}}{\partial r^{2}}+ \frac{2}{r}\frac{\partial C_{k}}{\partial r} - \frac{k(k+1)}{r^2}C_{k}, \\[2ex]
    \left.C_{k}\right|_{r=a=1} = \delta_{0k} ,\quad
    \left. C_{k}\right|_{r=\infty} = 0.
    \label{eq:spectral_r}
\end{split}
\end{align}
Here, the coefficients $E_{mnk}$ and $F_{mnk}$ are obtained by projection, using the orthogonality property of the Legendre polynomials,
\begin{equation} \label{eq: EFmnk}
\begin{split}
    E_{mnk} = \frac{2k+1}{2}\int_{-1}^{1}P_{n}P_{m}P_{k}d\mu,\qquad 
    F_{mnk} = \frac{(2k+1)}{2n(n+1)}\int_{-1}^{1}(1-\mu^{2})P'_{n}P'_{m}P_{k}d\mu.
\end{split}
\end{equation}
The terms $f_{nr}$ and $f_{n\theta}$ are the $r$ components of the flow field
\begin{equation}
\begin{split}
& \text{Sessile ciliated sphere:}  f_{nr} = \left(\frac{1}{{r}^{n+2}}-\frac{1}{{r}^{n}}\right),\quad
    f_{n\theta} = \left(\frac{n}{{r}^{n+2}}-\frac{n-2}{{r}^{n}}\right), \\
& \text{Motile ciliated sphere:} \; f_{1r} = \frac{2}{3}\left(-1 + \frac{1}{{r}^3}\right),\quad
    f_{1\theta} = \frac{2}{3}\left(2 + \frac{1}{{r}^3}\right),\\    
&  \qquad \qquad \qquad \qquad \qquad  f_{nr} = \left(\frac{1}{{r}^{n+2}}-\frac{1}{{r}^{n}}\right)\;(n\geq2),\quad
    f_{n\theta} = \left(\frac{n}{{r}^{n+2}}-\frac{n-2}{{r}^{n}}\right)\;(n\geq2), 
\end{split}
\label{fnrtheta}
\end{equation}

In the numerical calculation, we truncate the number of modes in the expansion  \eqref{eq:c_expand} of the concentration field to account for a finite number $M$ 
of modes.
To reach far-field boundary condition, we used a non-uniform radial mesh such that the grid is denser near the sphere surface and more sparse in the far-field. Specifically, we used the exponential function $r = e^{s(\zeta)}$, where $\zeta\in[0,1]$ and $s(\zeta)= w_1\zeta^3 + w_2\zeta^2 + w_3\zeta $ is a third order polynomial of $\zeta$ with constants $w_1, w_2, w_3$ chosen for getting converged results. To express~\eqref{eq:spectral_r} in terms of the transformed variable $\zeta$, we used the chain rule 
\begin{equation}
\begin{split}
    \frac{d C_k }{d r} = \frac{d C_k}{d\zeta}\frac{d\zeta}{d r},\quad
    \frac{d^{2} C_k}{d r^{2}} = \left(\frac{d\zeta}{d r}\right)^{2}\frac{d^{2} C_k }{d \zeta^{2}} + \frac{d^{2}\zeta}{d r^{2}}\frac{d C_k}{d \zeta}.
\end{split}
\end{equation}
Considering $N$ velocity modes and $M$  concentration modes, the differential equations in~\eqref{eq:spectral_r} can be rewritten as 
\begin{align}
    \begin{split}
      {\rm Pe}\sum_{n=1}^{N}\sum_{m=0}^{M}B_{n} \left(E_{mnk}f_{nr}\frac{dC_m}{d\zeta}\frac{d \zeta}{dr} - F_{mnk}f_{n\theta} \frac{C_{m}}{r}\right)\\
      -(\frac{d\zeta}{d r})^{2}\frac{d^2C_k}{d\zeta^2} - \frac{d^{2}\zeta}{d r^{2}} \frac{dC_k}{d\zeta}- \frac{2}{r}\frac{d\zeta}{d r}\frac{dC_k}{d\zeta}
      + \frac{k(k+1)}{r^2}C_k=0 .
    \end{split}
\end{align}
We discretized the spatial derivatives using the center difference scheme
\begin{equation}
\begin{split}
    \frac{dC_{m,j}}{d\zeta} = \frac{C_{m,j+1} - C_{m,j-1}}{2\triangle\zeta},\quad
    \frac{d^2C_{m,j}}{d\zeta^2} = \frac{C_{m,j+1} - 2C_{m,j} + C_{m,j-1}}{(\triangle\zeta)^2}.
    \label{eq:discrete}
\end{split}
\end{equation}

We computed the concentration field and Sherwood number for various Péclet number and tested the convergence of the results as a function of mesh size $\triangle\zeta$, computational domain $\mathcal{R}$, and the number of modes $M$ in the concentration expansion at Pe = 100. 
To test the effect of mesh size on the convergence of our simulations, we fixed the number of modes $M$ and computational domain size $\mathcal{R}$, and varied the mesh size as $\delta\zeta = [\frac{1}{100}, \frac{1}{200}, \frac{1}{400}, \frac{1}{800}]$. We computed the relative error of Sherwood number as a function of mesh size $\delta\zeta$; the results converged as the mesh size got smaller. 
Consistent with the second order accuracy of the discretization~\eqref{eq:discrete}, we obtained a convergence rate close to 2.
When testing the convergence as a function of the number of modes $M$ in the concentration expansion and the computational domain size $\mathcal{R}$, we found that a higher number of concentration modes and larger computational domain size are needed for higher P\'{e}clet numbers. 
Also, because the concentration field becomes more front and back asymmetric as P\'{e}clet increases, a denser mesh is required to capture the rapid change near the surface. 

\section{Optimization Method }
\label{append4}
To search for optimal surface motions that maximize feeding in the sessile  sphere model, we considered an optimization method based on variational analysis and steepest ascent \cite{michelin2011}.
The problem consists of a PDE-constrained optimization problem, where the goal is to find optimal $B_n$ that maximize the Sherwood number, subject to the concentration field $c$ satisfying the advection-diffusion equation and surface velocity satisfying the constant energy constraint,
\begin{equation}
\begin{split}
    & \underset{B_n, c}{\rm max}\; {\rm Sh}(B_n, c), \\ 
    & {\rm subject\; to}\quad \mathcal{L}[c]=0\quad \textrm{and} \quad
    \sum_{n=1}^{N} \frac{2B_n^2}{n(n+1)} = 1.
\end{split}
\end{equation}
Here, the linear operator $\mathcal{L}= {\rm Pe}\mathbf{u}\cdot\nabla - \nabla^2$ is that of the  advection-diffusion equation along with the corresponding boundary conditions. 

We use a variational approach to derive an adjoint system of equation. Given a surface motion, we consider small variations $\delta B_n$ in the coefficients associated with each mode. The corresponding small variations in the velocity field and concentration field are given by $\delta\mathbf{u}$ and $\delta c$, and result in a variation in Sh number \ref{eq:Sh}
\begin{align}
     \delta {\rm Sh} = -\frac{1}{4\pi}\int_S \nabla\delta c\cdot\mathbf{n}dS.
     \label{eq:dsh}
\end{align}
The concentration variation $\delta c$ must satisfy
\begin{align}
    {\rm Pe} (\mathbf{u}+\delta\mathbf{u})\cdot\nabla (c+\delta c) = \nabla^2 (c+\delta c), \quad (c+\delta c)(r=1) = 1, \quad (c+\delta c)(r\rightarrow\infty) = 0.
\end{align}
Subtracting the PDE for $c$ and keeping the leading order in $\delta c$, we arrive at
\begin{align}
    {\rm Pe} (\delta\mathbf{u}\cdot\nabla c + \mathbf{u}\cdot\nabla\delta c) = \nabla^2 \delta c,\quad \delta c(r=1) = 0, \quad \delta c(r\rightarrow\infty) = 0.
\end{align}
Multiplying the above equation with a test function $g(r,\mu)$ and integrating over the entire fluid domain, we get
\begin{align}
    {\rm Pe} \int_V(g\delta\mathbf{u}\cdot\nabla c + g\mathbf{u}\cdot\nabla\delta c)dV =  \int_V g\nabla^2 \delta c \;dV.
\end{align}
Using integration by parts and standard vector calculus identities, together with the appropriate boundary conditions and continuity property of the fluid, we obtain
\begin{align}
    -{\rm Pe}\int_V c\delta\mathbf{u}\cdot\nabla g\; dV - \int_S (g\nabla\delta c)\cdot \mathbf{n}dS = \int_V \delta c ({\rm Pe} \mathbf{u}\cdot\nabla g + \nabla^2g) dV,
\end{align}
where $dS$ is the surface element of the sphere with inward unit normal ${\mathbf{n}} = -\mathbf{e}_r$. Following a standard argument, we get that the test function $g$ must satisfy the following partial differential equation and boundary conditions 
\begin{align}
    {\rm Pe} \mathbf{u}\cdot\nabla g + \nabla^2g = 0, \quad g(r=1) = 1, \quad g(r\rightarrow\infty)=0,
\end{align}
and the consistency equation 
\begin{align}
    \int_S \nabla\delta c\cdot\mathbf{n}dS =-{\rm Pe}\int_V c\delta\mathbf{u}\cdot\nabla g\; dV.
    \label{eq:consistency}
\end{align}
From~\eqref{eq:dsh} and~\eqref{eq:consistency}, we get that the variation in Sh number is
given by
\begin{align}\label{eq:delta_J}
    \delta {\rm Sh} = -\frac{1}{4\pi}\int_S \nabla\delta c\cdot\mathbf{n}dS = -\frac{\rm Pe}{4\pi}\int_V c\delta\mathbf{u}\cdot\nabla g\; dV
    = -\frac{\rm Pe}{2}\int_1^\infty\int_{-1}^1 c(\delta\mathbf{u}\cdot\nabla g)\;r^2 dr d\mu .
\end{align}
We expand the test function $g(r,\mu)$ in terms of Legendre polynomials in $\mu$,
\begin{align}
    g(r,\mu) = \sum_{m=0}^{\infty} G_m(r)P_m(\mu),
\end{align}
and substitute back into~\eqref{eq:delta_J} with the perturbation from surface velocity $\delta \mathbf{u}(r=1) = (\delta B_n) V_n \mathbf{e}_\theta$. We arrive at  an expression for the gradient of nutrient uptake at each mode,
\begin{align} \label{eq:Sh_gradient}
    \frac{\delta {\rm Sh}}{\delta B_n} = -{\rm Pe}\sum_{m=0}^\infty\sum_{k=0}^\infty \frac{1}{{2k+1}}\int_1^\infty \left[ C_kf_{nr}G'_mE_{mnk} 
    - \frac{C_kG_mf_{n\theta}}{r}F_{mnk} \right]r^2 dr.
\end{align}
We now consider a finite number of velocity modes and express the input surface velocity as $\beta_n V_n(\mu)$, where $\beta_n = \sqrt{\frac{2}{n(n+1)}}B_n$. The weighted coefficients $\beta_n$ associated with each velocity mode must satisfy the constraint on the energy dissipation rate, that is, $\sum_n \beta_n^2 = 1$. 

Starting from an initial vector $\boldsymbol{\beta}^{0} = (\beta_1, \beta_2, \ldots, \beta_n,\ldots)$ of weighted coefficients, our goal is to find the optimal value of $\boldsymbol{\beta}$ that simultaneously maximize Sh and satisfy the constraint 
$ \| \boldsymbol{\beta}^{(j)} \| = 1$ at each iteration $j$ in the optimization process.
Thus, to get the value of $\boldsymbol{\beta}^{(j)}$ at subsequent iterations $j>0$, we project the feeding gradient onto the constraint space $ \| \boldsymbol{\beta}^{(j)} \| = 1$  using the linear projection $(I - \boldsymbol{\beta}^{(j)}\otimes \boldsymbol{\beta}^{(j)})$, where $I$ is the identity matrix. That is,
the steepest ascent direction of ${\rm Sh}$ with respect to weighted coefficients $\beta_n$ at the $j^{th}$ iteration is given by
\begin{align}
    \nabla_d {\rm Sh} = \nabla_\beta {\rm Sh} - (\boldsymbol{\beta}^{(j)}\cdot\nabla_{\beta}{\rm Sh})\boldsymbol{\beta}^{(j)}.
\end{align} 
The optimization process consists of updating $\boldsymbol{\beta}^{(j+1)}$ following the gradient ascent direction, where $\alpha$ is step size that can be adjusted in each iteration. 
\begin{align}
    \boldsymbol{\beta}^{(j+1)} = \frac{ \boldsymbol{\beta}^{(j)} + \alpha \nabla_d {\rm Sh}}{\|\boldsymbol{\beta}^{(j)} + \alpha \nabla_d {\rm Sh}\|}.
\end{align}


\renewcommand\thefigure{C.\arabic{figure}}
\setcounter{figure}{0}

\end{document}